\documentclass[epj,referee]{svjour}
\usepackage{graphics}
\begin{document}
\title{Towards feedback control of entanglement}
\author{Stefano Mancini\inst{1} \and Jin Wang\inst{2}
}                     
%
%
\institute{Dipartimento di Fisica, Universit\`a di Camerino,
62032 Camerino, Italy
\and
Department of Physics and Astronomy,
University of Nebraska-Lincoln, 65588 Lincoln NE
}
\date{Received: date / Revised version: date}
%
\abstract{
We provide a model to investigate feedback control of entanglement.
It consists of two distant (two-level) atoms
which interact through a radiation field
and becomes entangled.
We then show the possibility to stabilize such entanglement against
atomic decay by means of a feedback action.
\PACS{
      {03.67.Mn}{Entanglement manipulation}   \and
      {42.50.Lc}{Quantum fluctuations}
     } 
} 
\maketitle
\section{Introduction}
\label{introduction}

Over the last decade, the rapid development of quantum technology
has led to the possibility of continuously monitoring an individual
quantum system with very low noise and manipulating it on its typical
evolution time scale \cite{mab99}.
It is therefore natural to consider the possibility of controlling
individual quantum systems in real time by using feedback.
A theory of quantum-limited feedback has been introduced by Wiseman and
Milburn \cite{wismil93,wis94}.
Among recent developments we mention
the feedback stabilization of the state of a two level atom (single
qubit) against amplitude damping \cite{wanwis01}.

Because of the relevant role played by entanglement in quantum
processes, it would be straightforward
to also consider its feedback control.
Here we extend the basic idea of Ref.\cite{wanwis01}
to a recently proposed model \cite{manbos01} consisting of
two distant (two-level) atoms (two qubit)
which interact through a radiation field
and becomes entangled.
We then show the possibility to stabilize such entanglement against
atomic decay by means of a feedback action.

\section{The Model}
\label{model}

\begin{figure}
\begin{center}
\resizebox{0.85\columnwidth}{!}{
  \includegraphics{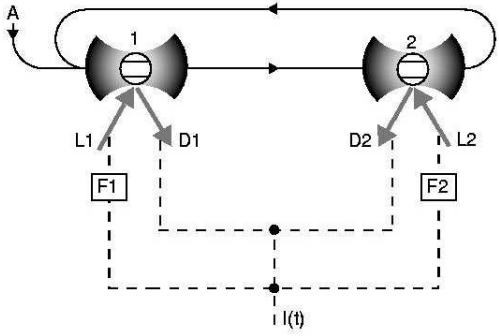}
}
\end{center}
\caption{ Schematic description of the considered set-up. Two
distinct cavites, each containing a two-level atom ($1$ and $2$
respectively), are connected via radiation fields (solid lines). A
coherent input of amplitude $A$ is provided in one of them.
Furthermore, $L1$, $L2$ and $D1$, $D2$ represent local operations,
namely driving fields and homodyne detection respectively. $I$ is
the current arising from local measurements and $F1$, $F2$
indicate the consequent local feedback actions (dashed lines). }
\label{fig1}
\end{figure}

We consider a very simple model consisting of two two-level atoms,
$1$ and $2$, placed in distant cavities and interacting through
a radiation field in a dispersive way.
The two cavities are arranged in a cascade-like
configuration such that,
given a coherent input field with amplitude $A$ in one of them,
the output of each cavity enters the other
as depicted in Fig.\ref{fig1}.
Then, it is possible to show \cite{manbos01},
after eliminating the radiation fields, that
the effective interaction Hamiltonian for the internal degrees
of the two atoms becomes of Ising type, namely
\begin{equation}\label{eq:Heff}
H_{int}=2J\sigma_{1}^{(z)}\sigma_{2}^{(z)}\,,
\end{equation}
where $\sigma_{1,2}^{(x,y,z)}$
indicate the usual Pauli operators. Hereafter we shall also use
$\sigma_{1,2}\equiv(\sigma_{1,2}^{(x)}+i\sigma_{1,2}^{(y)})/2$.
The spin-spin coupling constant $J$ scales as
radiation pressure $|A|^{2}$ and goes to
zero for negligible cavity detuning \cite{manbos01}.

To get entanglement in an Ising model,
it is necessary to break its symmetry \cite{gunetal01}.
To this end, we consider local
laser fields applied to each atom
(L1 and L2 of Fig.\ref{fig1}) such that
a local Hamiltonian $H_{drive}$ given by
\begin{equation}\label{eq:Hlocal}
H_{drive}=\alpha \sigma_1^{(y)} + \alpha \sigma_2^{(y)},
\qquad (\alpha\in{\bf R})
\end{equation}
acts in addition to $H_{int}$. Thus, the
total Hamiltonian of the system results
\begin{equation}\label{eq:Htot}
H_{tot} = H_{drive} + H_{int}.
\end{equation}

Let us introduce the ground and excite atomic states $|g\rangle_{1,2}$,
$|e\rangle_{1,2}$ as eigenvectors of $\sigma_{1,2}^{(z)}$ with
$-1$ and $+1$ eigenvalues respectively, and $\eta\equiv\alpha/J$.
Then, the eigenvectors of the  Hamiltonian $H_{tot}$ read
\begin{eqnarray}\label{egvec}
|\psi_{1}\rangle &=&
\frac{\eta}{2\sqrt{1+\eta^{2}+\sqrt{1+\eta^{2}}}}
\left(|g\rangle_{1}|g\rangle_{2}-|e\rangle_{1}|e\rangle_{2}\right)
\nonumber\\
&& +i \frac{1+\sqrt{1+\eta^{2}}}
{2\sqrt{1+\eta^{2}+\sqrt{1+\eta^{2}}}}
\left(|e\rangle_{1}|g\rangle_{2}+|g\rangle_{1}|e\rangle_{2}\right)\,,
\nonumber\\
|\psi_{2}\rangle&=&\frac{1}{\sqrt{2}}\left(
|e\rangle_{1}|g\rangle_{2}-|g\rangle_{1}|e\rangle_{2}\right)\,,
\nonumber\\
|\psi_{3}\rangle&=&\frac{1}{\sqrt{2}}\left(
|g\rangle_{1}|g\rangle_{2}+|e\rangle_{1}|e\rangle_{2}\right)\,,
\nonumber\\
|\psi_{4}\rangle &=&
\frac{\eta}{2\sqrt{1+\eta^{2}-\sqrt{1+\eta^{2}}}}
\left(|g\rangle_{1}|g\rangle_{2}-|e\rangle_{1}|e\rangle_{2}\right)
\nonumber\\
&& +i \frac{1-\sqrt{1+\eta^{2}}}
{2\sqrt{1+\eta^{2}-\sqrt{1+\eta^{2}}}}
\left(|e\rangle_{1}|g\rangle_{2}+|g\rangle_{1}|e\rangle_{2}\right)\,,
\nonumber\\
\end{eqnarray}
with eigenvalues
$E_{1}=-2\sqrt{\alpha^{2}+J^{2}}$,
$E_{2}=-2 J$, $E_{3}=2 J$ and
$E_{4}=2\sqrt{\alpha^{2}+J^{2}}$.

It is reasonable to consider as initial state of the two atoms
the ground state $|g\rangle_{1}|g\rangle_{2}$; then
we can expand it
over the eigenstates basis (\ref{egvec}) as
\begin{equation}\label{Psi0}
|\Psi(0)\rangle\equiv |g\rangle_{1}|g\rangle_{2}=\sum_{j=1}^{4}
C_{j}|\psi_{j}\rangle\,,
\end{equation}
with
\begin{eqnarray}
C_{1}&=&-\frac{\left(1-\sqrt{1+\eta^{2}}\right)
\sqrt{1+\eta^{2}+\sqrt{1+\eta^{2}}}}
{2\eta\sqrt{1+\eta^{2}}}\,,
\nonumber\\
C_{2}&=&0\,,
\nonumber\\
C_{3}&=&\frac{1}{\sqrt{2}}\,,
\nonumber\\
C_{4}&=&\frac{\left(1+\sqrt{1+\eta^{2}}\right)
\sqrt{1+\eta^{2}-\sqrt{1+\eta^{2}}}}
{2\eta\sqrt{1+\eta^{2}}}\,.
\end{eqnarray}

\section{System dynamics}
\label{sysdyn}

The evolution of the state (\ref{Psi0})
under $H_{tot}$ gives
\begin{eqnarray}\label{Psit}
|\Psi(t)\rangle&=&C_{1}e^{2i\tau\sqrt{1+\eta^{2}}}|\psi_{1}\rangle
\nonumber\\
&&+C_{2}e^{-2i\tau}|\psi_{2}\rangle
\nonumber\\
&&+C_{4}e^{-2i\tau\sqrt{1+\eta^{2}}}|\psi_{4}\rangle\,,
\end{eqnarray}
where we have introduced the scaled time $\tau=J t$.

In Ref.\cite{wanwis01} it was shown that homodyne measurement of the
light scattered by an atom allows indirect measurement of its spin flip
operators. Then, let us consider,
such type of local measurements so that after combining homodyne currents,
the total current $I(t)$ carries out information about the observable
${\cal O}\equiv
\sigma_{1}^{(x)}-\sigma_{2}^{(x)}$.
Its variance over the state (\ref{Psit}) is
\begin{eqnarray}
\label{marker}
&&\langle \Psi(t)|{\cal O}^{2}
|\Psi(t)\rangle
-\langle \Psi(t)|{\cal O}
|\Psi(t)\rangle^{2}
\nonumber\\
&&=2-\frac{\eta^{2}}{1+\eta^{2}}
\left[1-\cos\left(4\tau\sqrt{1+\eta^{2}}\right)\right]\,.
\end{eqnarray}
Notice that this quantity being strictly less than $2$
at almost any time, shows the presence of correlations for the state (\ref{Psit}).
On the other hand, in Ref.\cite{manbos01} it has been shown
that the state (\ref{Psit}) exhibits entanglement at
almost any time. We are thus led to ascribe the correlations of Eq.(\ref{marker}) to the presence of entanglement in Eq.(\ref{Psit}), though this would not generally true.
Then, we are going to consider the quantity ${\cal O}$ as a ``marker" of
entanglement  while characterizing the open system dynamics.

When we include the effect of
spontaneous atomic decay at rate $\gamma$, the dynamics of the two
distant atoms is described by the master equation
\begin{eqnarray}\label{menofb}
\dot{\rho} &=& -i\left[H_{tot},\rho\right]
+{\bf D}\left[\sigma_{1}\right]\rho
+{\bf D}\left[\sigma_{2}\right]\rho
\nonumber\\
&\equiv&-i\left[H_{tot},\rho\right]
+{\bf D}\left[c_{+}\right]\rho
+{\bf D}\left[c_{-}\right]\rho
\,,
\end{eqnarray}
where $c_{\pm}=(\sigma_{1}\pm\sigma_{2})/\sqrt{2}$ and
${\bf D}$ is the Lindblad decoherence superoperator, i.e.
${\bf D}[a]b\equiv aba^{\dag}-a^{\dag}ab/2-ba^{\dag}a/2$.
The following replacements $J/\gamma\to J$,
$\alpha/\gamma\to \alpha$, $\gamma t\to t$ have been made
deriving Eq.(\ref{menofb}).

The steady state solution of Eq.(\ref{menofb})
can be easily found by writing the density
operator in a matrix form, in the basis
$\{|e\rangle_{1}|e\rangle_{2},
|g\rangle_{1}|e\rangle_{2},
|e\rangle_{1}|g\rangle_{2},
|g\rangle_{1}|g\rangle_{2}\}$, as
\begin{equation}\label{rhoexp}
\rho_{ss}=
\left(\begin{array}{cccc}
{\cal A}&{\cal B}_{1}+i{\cal B}_{2}&
{\cal C}_{1}+i{\cal C}_{2}&{\cal D}_{1}+i{\cal D}_{2}
\\
{\cal B}_{1}-i{\cal B}_{2}&{\cal E}&
{\cal F}_{1}+i{\cal F}_{2}&{\cal G}_{1}+i{\cal G}_{2}
\\
{\cal C}_{1}-i{\cal C}_{2}&{\cal F}_{1}-i{\cal F}_{2}&
{\cal H}&{\cal I}_{1}+i{\cal I}_{2}
\\
{\cal D}_{1}-i{\cal D}_{2}&{\cal G}_{1}-i{\cal D}_{2}&
{\cal I}_{1}-i{\cal I}_{2}&{\cal L}
\end{array}\right)\,,
\end{equation}
while the matrix representation of the other operators
(in the same basis) comes from
\begin{equation}
\sigma_{1}=
\left(\begin{array}{cccc}
0&0&0&0
\\
1&0&0&0
\\
0&0&0&0
\\
0&0&1&0
\end{array}\right)\,,
\quad
\sigma_{2}=
\left(\begin{array}{cccc}
0&0&0&0
\\
0&0&0&0
\\
1&0&0&0
\\
0&1&0&0
\end{array}\right)\,.
\end{equation}
By inserting these matrices in the r.h.s. of Eq.(\ref{menofb})
and equating to $0$ at l.h.s. we are left with a set of $16$ linear equations
from which we can calculate (together with the normalization
condition ${\rm tr}(\rho)=1$) all the real coefficients
of the matrix (\ref{rhoexp}), namely
\begin{eqnarray}\label{rhoss}
{\cal A}&=&\frac{1}{\Xi}
16 \alpha^{4}, \nonumber\\
{\cal B}_{1}&=&-\frac{1}{\Xi}8 \alpha^{3},
\qquad {\cal B}_{2}=0, \nonumber\\
{\cal C}_{1}&=&-\frac{1}{\Xi}8 \alpha^{3},
\qquad {\cal C}_{2}=0, \nonumber\\
{\cal D}_{1}&=& \frac{1}{\Xi} 4\alpha^{2},
\qquad {\cal D}_{2}=\frac{1}{\Xi}16\alpha^{2} J,\nonumber\\
{\cal E}&=&\frac{1}{\Xi}\left(16 \alpha^{4}+4\alpha^{2}\right),
\nonumber\\
{\cal F}_{1}&=&\frac{1}{\Xi}4\alpha^{2},
\qquad {\cal F}_{2}=0, \nonumber\\
{\cal G}_{1}&=&-\frac{1}{\Xi}2\alpha(4\alpha^{2}+1),
\qquad {\cal G}_{2}=-\frac{1}{\Xi}8\alpha J, \nonumber\\
{\cal H}&=&\frac{1}{\Xi}\left(16 \alpha^{4}+4\alpha^{2}\right),
\nonumber\\
{\cal I}_{1}&=&-\frac{1}{\Xi}2\alpha(4\alpha^{2}+1),
\qquad {\cal I}_{2}=-\frac{1}{\Xi}8\alpha J, \nonumber\\
{\cal L}&=&\frac{1}{\Xi}\left(16
\alpha^{4}+8\alpha^{2}+1+16J^{2}\right),
\end{eqnarray}
with
\begin{equation}
\Xi=64\alpha^{4}+16\alpha^{2}+1+16J^{2}\,.
\end{equation}

\section{Stationary entanglement}
\label{entanglement}

One can use the concurrence as measure of the degree of
entanglement between two qubit described by density operator $\rho$ \cite{woo98}.
It is defined as
\begin{equation}
C(\rho)=\max\left\{0,\xi_{1}-\xi_{2}
-\xi_{3}-\xi_{4}\right\}
\end{equation}
where $\xi_{i}$'s are, in decreasing order, the nonnegative square
roots of the moduli of the eigenvalues of the non-hermitian matrix
$\rho\tilde\rho$.
Here $\tilde\rho$ is the matrix given by
\begin{equation}
\tilde\rho\equiv\left(\sigma_{1}^{(y)}\otimes\sigma_{2}^{(y)}\right)
\rho^{*}\left(\sigma_{1}^{(y)}\otimes\sigma_{2}^{(y)}\right)\,,
\end{equation}
where $\rho^{*}$ denotes
the complex conjugate.

\begin{figure}
\begin{center}
\resizebox{1.0\columnwidth}{!}{
  \includegraphics{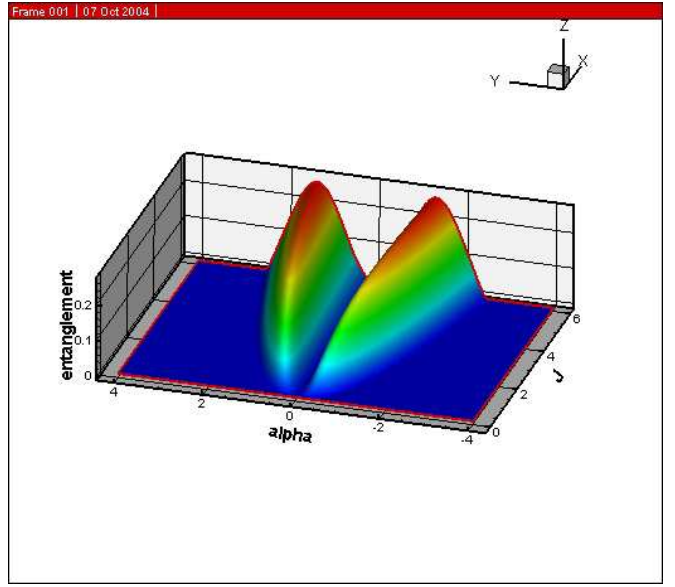}
}
\end{center}
\caption{Concurrence $C_0$ of the steady state plotted versus
the driving strength $\alpha$ and
the coupling constant $J$.
}
\label{fig2}
\end{figure}

The stationary state concurrence $C_0\equiv C(\rho_{ss})$
is shown in Fig.\ref{fig2}.
It is clear that a relevant amount of entanglement persists
at steady state only for large values of the coupling constant,
i.e.  $J\gg1$, (when the original $J$ is much greater
than $\gamma$).

\section{Feedback action}
\label{feedback}

We can now think to stabilize the entanglement, i.e. to prevent its
degradation, by using a feedback action on the driving fields
(L1 and L2 of Fig.\ref{fig1})
accordingly to the measured quantity
${\cal O}$ which should reveal the status of
nonclassical correlations.
Then we act on the system with a local feedback operator
\begin{equation}
F\equiv\frac{\lambda}{\sqrt{2}}
\left(\sigma_{1}^{(y)}-\sigma_{2}^{(y)}\right)\,,
\end{equation}
where $\lambda$ represent the feedback strength
(already scaled by $\gamma$, i.e. $\lambda/\sqrt{\gamma}\to\lambda$).
The choice of $F$ is motivated by the fact that feedback
mediated by indirect (homodyne) measurement
requires, to squeeze the variance of
a variable (${\cal O}$), a  driving action
on the conjugate variable \cite{man00}.

The master equation (\ref{menofb}) then becomes \cite{wis94}
\begin{eqnarray}\label{me}
\dot{\rho} &=& -i\left[H_{tot},\rho\right]
+{\bf D}\left[c_{+}\right]\rho
\nonumber\\
&&+{\bf D}\left[c_{-}-iF\right]\rho
-\frac{i}{2}\left[c_{-}^{\dag}F+Fc_{-},\rho\right]\,.
\end{eqnarray}
In the above equation,
the feedback operator $F$ appears in the Hamiltonian term describing the driving effect,
as well as inside the decoherence superoperator accounting for quantum noise
carried back into the system from measurement.

\begin{figure}
\begin{center}
\resizebox{1.0\columnwidth}{!}{
  \includegraphics{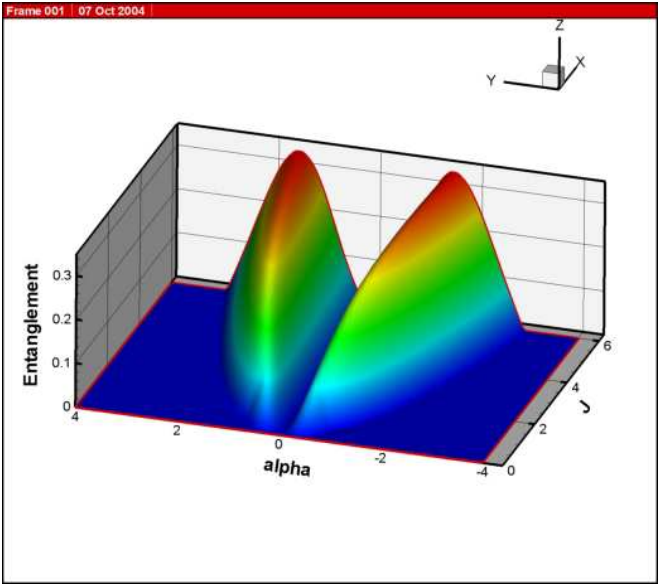}
}
\end{center}
\caption{Concurrence $C_{fb}$ of the steady state plotted versus
 the driving strength $\alpha$ and
the coupling constant
$J$ in presence of feedback action.
For each value of $\alpha$ and $J$, the feedback strength is chosen to be the optimal.}
\label{fig3}
\end{figure}

The master equation (\ref{me}) can be solved at steady state
with the same method of Eq.(\ref{menofb}), obtaining $\rho^{fb}_{ss}$.
However,  the analytical expression is quite cumbersome,
hence not reported at all.
The state $\rho^{fb}_{ss}$ allows us to (numerically) calculate its concurrence.
In particular, we have evaluate the quantity
\begin{equation}\label{Cfb}
C_{fb}\equiv\max_{\lambda\in{\bf R}}C(\rho^{fb}_{ss})\,,
\end{equation}
that is shown in Fig.\ref{fig3}.
We can see that feedback improves the available
entanglement with respect to previous case (Fig.\ref{fig2}).
Feedback seems especially powerful at small values of $J$
where entanglement was very fragile
(it somehow enforces the coupling effect).

\begin{figure}
\begin{center}
\resizebox{1.0\columnwidth}{!}{
  \includegraphics{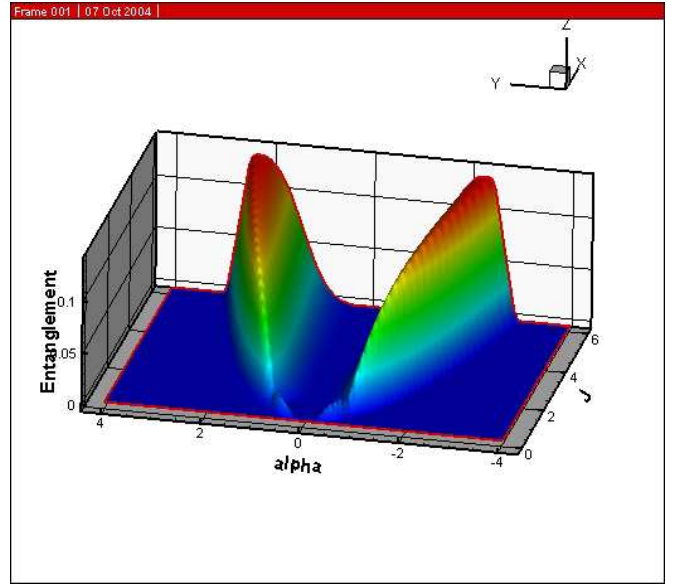}
}
\end{center}
\caption{Concurrences difference $C_{fb}-C_0$ plotted versus
the driving strength $\alpha$ and
the coupling constant
$J$.
}
\label{fig4}
\end{figure}

To better compare the results with and without feedback,
in  Fig.\ref{fig4} we have shown the difference $C_{fb}-C_0$.

\section{Conclusion}
\label{conclusion}

We have shown the possibility to improve the steady state entanglement
in an open quantum system by using a feedback action.
Although the improvement is not very high the above result
represents a proof of
principle about the possibility of
controlling entanglement through feedback.
A complementary possibility to
increase entanglement between atoms subject to joint measurements
with feedback has been then proposed \cite{wanwismil04}.

Since our method only relies on Local Operations
and Classical Communication (LOCC), what we have obtained is perhaps related to
\textit{entanglement purification} \cite{ben96}.

To improve the presented model one should find the best
entanglement witness \cite{ecketal03} to measure,
and then optimize the feedback action
(operator). This can be phrased in terms of a numerical optimization problem and
 is left for future work.
Moreover, since entanglement is a system state peculiarity,
other feedback procedures, like
state estimation based feedback \cite{doh99}, could be more powerful.

Summarizing, although  we have proved the possibility
of feedback control of entanglement,
its effectiveness remains difficult to quantify in
nonlinear systems (like that studied).
Probably,
investigations in linear systems would be more fruitful.

\section*{Acknowledgments}
The authors warmly thank H. M. Wiseman for insightful comments.

\end{document}